\documentclass[iop]{emulateapj}
\usepackage{apjfonts}

\newcommand{\SDSS}{SDSS\,1257+5428}

\slugcomment{Submitted to the Astrophysical Journal, 10 March 2010}
\begin{document}
\title{The (Double) White Dwarf Binary SDSS\,1257+5428}
\author{S. R. Kulkarni and M. H. van Kerkwijk\altaffilmark{1}}
\affil{Caltech Optical Observatories 249--17, California Institute of
  Technology, Pasadena, CA 91125} 
\altaffiltext{1}{On sabbatical leave from Department of Astronomy and
  Astrophysics, University of Toronto, 50 St. George Street, Toronto,
  ON M5S 3H4, Canada.} 

\begin{abstract}
  \SDSS\ is a white dwarf in a close orbit with a companion that has
  been suggested to be a neutron star.  If so, it hosts the closest
  known neutron star, and its existence implies a great abundance of
  similar systems and a rate of white-dwarf neutron-star mergers
  similar to that of the type Ia supernova rate.  Here, we present
  high signal-to-noise spectra of \SDSS, which confirm an independent
  finding that the system is in fact composed of two white dwarfs, one
  relatively cool and with low mass, and the other hotter and more
  massive.  With this, the demographics and merger rate are no longer
  puzzling (various factors combine to lower the latter by more than
  two orders of magnitude).  We show that the spectra are fit well
  with a combination of two hydrogen model atmospheres, as long as the
  lines of the higher-gravity component are broadened significantly
  relative to what is expected from just pressure broadening.
  Interpreting this additional broadening as due to rotation, the
  inferred spin period is short, about 1 minute.  Similarly rapid
  rotation is only seen in accreting white dwarfs that are magnetic;
  empirically, it appears that in non-magnetized white dwarfs,
  accreted angular momentum is lost by nova explosions before it can
  be transferred to the white dwarf.  This suggests that the massive
  white dwarf in \SDSS\ is magnetic as well, with $B\simeq10^5\,$G.
  Alternatively, the broadening seen in the spectral lines could be
  due to a stronger magnetic field, of $\sim\!10^6\,$G.  The two
  models can be distinguished by further observations.
\end{abstract}

\keywords{white dwarfs: general ---
white dwarfs: individual (\objectname{SDSS 1257+5428})}

\section{\SDSS}

An unexpected return from the Sloan Digital Sky Survey (SDSS;
\citealt{yaa+00}) has been the discovery of white dwarf binaries with
short periods, by \citet{bmt+09} and \citet{mbt+09}.  The massively
multiplexed SDSS spectroscopic observations consist of several
15-minute integrations, which are combined to yield the final spectra.
\citet{mbt+09} and \citet{bmt+09} took advantage of this approach and
looked for rapid changes in radial velocity across the individual
spectra.

The first finds are already quite interesting.
SDSS\,J143633.29+501026.8 (orbital period, $P_{\rm b}=$1.15\,hr) and
SDSS\,J105353.89+520031.0 ($P_{\rm b}=0.96$\,hr) are both low mass
($\sim$0.3\,M$_\odot$) DA white dwarfs orbiting an unseen
companion. For both systems, the measured mass function can be
reasonably explained by invoking a white dwarf for the
secondary.\footnote{We refer to the detected DA white dwarf as the
  photometric primary (or ``primary'') and the unseen/fainter
  companion as the photometric secondary (``secondary'').} The orbital
periods are so short that the systems are expected to merge within
a Hubble time. These two binaries increased the toll of such
interesting double degenerates to seven \citep{mbt+09}.  The number is
likely to increase further; e.g., from a targeted search of low-mass
white dwarfs found in SDSS, \citet{kba+09} find four short-period
binaries (including the two from \citealt{mbt+09}).  These authors
also discuss in detail the fate of these systems: since the combined
masses are likely below the Chandrasekhar mass, most will not explode
as type Ia supernovae, but rather become AM CVn systems or R CrB stars.

SDSS\,J125733.63+542850.5 (\SDSS\ hereafter) consists of a DA white
dwarf primary in an orbit with period $P_{\rm
  b}=4.6$\,hr and no measurable eccentricity.  The inferred companion
mass (as constrained by the spectroscopically inferred white-dwarf
mass and the mass function) is above the Chandrasekhar mass (in a
probabilistic sense).  If so, the companion is either a neutron star
or a black hole, and, at the estimated distance, $d\approx 50$\,pc, it
would be the nearest ultra-compact object known; see \citet{bmt+09}.
%  Next, the expected coalescence time is
%  less than a billion years.  Thus the outcome is expected to be a
%  bright cosmic explosion of an entirely new sort but with a frequency
%  comparable to that of supernovae of type Ia \citep{tks09}.

The importance of the proximity of \SDSS\ becomes apparent when we
consider the  distances to the nearest members of different sub-classes
of neutron stars (see Table~\ref{tab:NearestNS}): the nearest
millisecond binary pulsar (a binary consisting of a low-mass white
dwarf and a pulsar and in a circular orbit), PSR\,J0437$-$4715; the
thermally X-ray emitting middle aged neutron star, RX~J1856.5$-$3754;
the nearest long period pulsar, PSR\,J2144$-$3933; the nearest
ordinary pulsars, PSR\,J0108$-$1431 and PSR\,0950+08; the nearest
young pulsar, Vela and the nearest $\gamma$-ray pulsar, Geminga.
(For the ordinary pulsars, accounting for the usual beaming factor
of 0.1, results in a distance of about 130\,pc.)  The distance
$d_{\mathrm{NS}}\sim 150\,$pc for the nearest neutron star(s) is
in accord with the global demographics of neutron stars as well as
demographics of specific classes of neutron stars (based on birth
rates).

Specifically for \SDSS, if its unseen companion is truly a neutron
star then, given the short period, the system would arguably be a
binary millisecond pulsar.  The local surface density of pulsars and
millisecond pulsars is about 30\,kpc$^{-2}$ \citep{lml+98}.  For
millisecond pulsars the local density is between 30--45\,kpc$^{-3}$
\citep{cc97}. The expected distance to the nearest millisecond pulsar
is thus 175--200\,pc, consistent with the observations but
inconsistent with the proximity of \SDSS.

\begin{deluxetable}{lll}
\tablecaption{The Nearest Neutron Stars.\label{tab:NearestNS}}
\tablewidth{\hsize}
\tablehead{
&\colhead{Distance}&\\
\colhead{Name}&\colhead{(pc)}&\colhead{Ref.}}
\startdata
PSRJ0437$-$4715 & $156.3\pm1.3$      & \citealt{dvt+08}\\
RXJ1856.5$-$3754 & $161^{+18}_{-14}$ & \citealt{vkk07}\\
PSRJ2144$-$3933  & $165^{+17}_{-14}$   & \citealt{dtb+09}\\
PSRJ0108$-$1431  & $240^{+124}_{-61}$ & \citealt{dtb+09}\\
PSR0950+08           & $260^{+58}_{-5}$   & \citealt{bbg+02}\\
Vela Pulsar                & $286^{+19}_{-17}$  & \citealt{dlr+03}\\
Geminga                    & $250^{+120}_{-62}$ & \citealt{fwa07}
\enddata
\end{deluxetable}

%The local density  of objects similar
% to \SDSS\  relative to that of low-mass binary millisecond pulsars
% is very large, $(d_{\mathrm{NS}}/D)^3\approx 27$. This inference
%flies in the face of all we know about the demography of neutron
%stars.  We believe that a critical re-examination of the nature of
%the unseen companion of \SDSS\ is urgently needed.

Thus, a neutron-star companion to \SDSS\ would be very puzzling,
unless one assumes that its proximity to the solar system is a
statistical fluke.  Motivated by this, as well as by the alarming
implications of a paper by \cite{tks09}, we first reconsider the
case for a neutron-star companion.  A reading of the \cite{bmt+09}
paper shows that it rests primarily on the high mass for the primary,
with kinematics offered as a supporting argument.  We revisit these
two issues in the next two sections, and then present observations
showing that the system in fact hosts two white dwarfs.

Before proceeding, we note that while writing up our results, a
preprint by \citet{mgs+10} appeared, which presented lower-quality but
much more numerous spectra, from which the presence of a second
component is also evident.  These authors discuss in detail the
preceding and future evolution of this binary.  In this paper, we
focus on the demographics, and on obtaining more accurate constraints
on the white dwarfs in the system from our higher-quality spectra.

\section{Revisiting the Mass of the Companion}
\label{sec:RevisitingMass}

\cite{bmt+09} present radial-velocity variations of \SDSS. The
velocity data appear to be of high quality and so we accept the two
inferences: orbital period, $P_{\rm b}=4.555\,$hr and a velocity semi-major
amplitude of the primary, $K_1=322.7\pm 6.3\,$km\,s$^{-1}$.  The
uncertainty in $P$ is virtually negligible.  The mass function is thus
\begin{equation}
	\mathcal{M}_2 =\frac{M_2^3\sin^3 i}{(M_1+M_2)^2}
        =\frac{P_{\rm b}K_1^3}{2\pi G} = 0.66\pm 0.04\,M_\odot
\end{equation}
where the uncertainty of 5.9\% of $\mathcal{M}_2$ arises from cubing
$K_1$.  The minimum mass of the unseen companion is $\mathcal{M}_2$
(obtained by assuming that the system is seen edge-on and setting
$M_1=0$).

\citet{bmt+09} infer a mass for the primary based on fitting the
Balmer absorption lines to model atmospheres of DA white dwarfs and
find $M_1=0.92^{+0.28}_{-0.32}\,M_\odot$. The large uncertainty in the
inferred mass is due to the poor model fit to the data (more on this
issue below).

As noted by \citet{bmt+09}, the unseen companion is quite massive
(for a white dwarf) and exceeds the Chandrasekhar limit for
$M_1>0.64\,M_\odot$. There is an 8\% chance that $M_1$ is less than
this value and thus the above appears evidence for the conclusions of
\citet{bmt+09}.

One issue is that empirically it has been found that spectroscopic
masses inferred for cool white dwarfs ($T_\mathrm{eff}<12,000\,$K) are
larger than the true masses.  This has been a long-standing issue in
the field of white dwarfs, but with no resolution in sight; see
\citet{tbk+10} and \citet{kkk+09} for recent summaries.  Spectral
fitting yields $P=g/\kappa $, where $P$ is the gas pressure, $\kappa$
is the mass opacity, and $g$ is the local acceleration due to gravity.
It has been suspected that in the cool white dwarfs, which have
convective atmospheres, helium might be dredged up.  This would
pollute the atmosphere and decrease $\kappa$ (relative to a pure
hydrogen atmosphere).  This hypothesis would then explain why the
masses inferred spectroscopically are systematically higher for cool
DA white dwarfs.  Sensitive measurements in field white dwarfs,
however, have failed to detect the expected \ion{He}{1} features at
$5876$\,\AA\ \citep{tbk+10}.

Ignoring this very interesting and perplexing issue we proceed in an
empirical fashion.  Using the run of inferred mass as a function of
$T_{\mathrm{eff}}$ for DA white dwarfs, we derive a multiplicative
correction factor of 0.9 for the mass inferred from spectroscopic
modeling.  The resulting decrease in mass to $M_1=0.8\pm0.3$ also
decreases the mass estimate for the secondary and thus weakens the
case for a neutron-star companion.

Finally, and more importantly, the spectral fits to the data are poor.
This is acknowledged by \citet{bmt+09}, but it is worth stressing that
such poor fits are not routinely seen in DA white dwarfs (e.g., see
\citealt{tbd09}).  That typical fits are good can also be seen from the
small inferred uncertainties in $T_{\mathrm{eff}}$ and $\log g$.
Taking the SPY survey of high sensitivity and high spectral resolution
observations of white dwarfs as an example, and restricting ourselves
to the subset of cool DA white dwarfs ($T<12,000\,$K), we find that
the largest uncertainty in $\log g$ is 0.03 (corresponding to 7\%
uncertainty) and the median 0.01 (corresponding to 2.3\%).

Thus, the poor fit for the Balmer lines of \SDSS\ suggest that either
\SDSS\ has a strange pathology (in which case modeling of the Balmer
lines is fundamentally suspect) or that the companion is contributing
to the light.  If the latter, it might be responsible for some of the
wiggles seen in the continuum, or, if it is a DA white dwarf,
contribute to the observed Balmer lines.  From a more detailed look at
the spectral fits of \citet{bmt+09}, one sees that these poorly match
the narrow cores of lower Balmer lines and under-predict the higher
Balmer lines.  The presence of the higher Balmer lines is indicative
of lower gravity, since these lines are very sensitive to pressure
broadening (the size of a hydrogen atom is quadratically proportional
to the upper level of the transition).  Indeed, the fact that in the
spectrum presented by \cite{bmt+09}, the line at H10 has depth
comparable to that of H9 already informs us that $\log g<7$.  This
inference, combined with the wide wings shown by the lower Balmer
lines, supports the idea that the observed spectrum of \SDSS\ is
produced by more than one DA white dwarf.

\section{Revisiting the Kinematics}
\label{sec:RevisitingKinematics}

Pulsars, both single and binary, are noted for their large space
motion \citep{cc97}. The mean {3-D} speed of the low-mass millisecond
pulsars is 84\,km\,s$^{-1}$.  \citet{bmt+09} offer the kinematics of
\SDSS\ as a consistency argument for a neutron star secondary.

The proper motion of \SDSS\ is $0.049{\rm\,mas\,yr^{-1}}$.  At the
stated distance of 50\,pc the proper motion translates to $11\,{\rm
  km\,s^{-1}}$, which is not particularly distinguishable.
\citet{bmt+09}, from their velocities, find that the mean radial
velocity of the primary is $\gamma_1=-29\pm 5{\rm\,km\,s^{-1}}$.

A compilation of the mean radial velocities\footnote{For consistency,
  we restricted the analysis to the radial velocity of the photometric
  primary.}  of double degenerates listed by \citet{nnk+05} shows a
spread of $-36$ to $70{\rm\,km\,s^{-1}}$.  Correcting for the expected
gravitational redshift\footnote{Throughout this article we use the
  mass-radius fitting formula of \citet{n72} which provides a good
  approximation to the classical results of \citet{hs61}.  The
  gravitational or Einstein redshift, converted to velocity, is
  $\gamma_E=0.633\,M/R\,$km\,s$^{-1}$ where $M$ is the mass and $R$ is
  the radius.}  for the primary, using the masses as obtained from
spectroscopic modeling, one finds that the true radial velocities
for these 24 double degenerates lie between $-60$ and
$50{\rm\,km\,s^{-1}}$.

For the corrected spectroscopic mass of $0.8\pm0.3\,M_\odot$, the
gravitational redshift (converted to velocity) is
$56^{+58}_{-30}{\rm\,km\,s^{-1}}$.  The 1-$\sigma$ lower bound results
in $-55\pm5{\rm\,km\,s^{-1}}$ -- well within the expected kinematics
of double degenerates.

\begin{deluxetable}{lllllll}
\tablewidth{\hsize}
\tablecaption{Log of Observations and Radial Velocity
  Measurements\label{tab:KeckLog}}
\tablehead{
\colhead{Epoch}&
\colhead{Slit}&
\colhead{Cam}&
\colhead{$\tau$}&
\colhead{MJD}&
\colhead{$\phi$}&
\colhead{$v$}\\
\colhead{(1)}&\colhead{(2)}&\colhead{(3)}&\colhead{(4)}&
\colhead{(5)}&\colhead{(6)}&\colhead{(7)}}
\startdata
12:13& 0.7& B&  780 & 0.51660& 0.86\\ 
12:14& 0.7& R&  340 & 0.51470& 0.85&      $-292\pm3$\\ 
12:21& 0.7& R&  340 & 0.51945& 0.87&      $-262\pm3$\\[0.8ex] 
12:28& 1.0& B&  780 & 0.52727& 0.92\\ 
12:28& 1.0& R&  340 & 0.52491& 0.90&      $-220\pm3$\\ 
12:35& 1.0& R&  340 & 0.52967& 0.93&      $-173\pm3$\\[0.8ex] 
12:42& 1.0& B&  780 & 0.53709& 0.97\\ 
12:43& 1.0& R&  340 & 0.53473& 0.96&      $-120\pm3$\\ 
12:49& 1.0& R&  340 & 0.53950& 0.98&\phn   $-71\pm3$
\enddata
\tablecomments{Col.\ (1): Exposure start time on 2010 February 11.
               Col.\ (2): Slit width in arcseconds.
               Col.\ (3): Camera (B: Blue, R: Red).
               Col.\ (4): Integration time in seconds.
               Col.\ (5): Barycentric mid-exposure time relative to MJD 55238.0.
               Col.\ (6): Orbital phase calculated using the ephemeris
               of \citet{mgs+10}, MJD $54845.67470(8)+0.18979154(9)E$,
               where phase 0 corresponds to inferior conjunction of the
               photometric primary.
               Col.\ (7): Radial velocity calibrated relative to sky
               lines, and corrected to the solar system barycenter.
               There may be  systematic effects larger than the formal
               uncertainties due to inaccuracies in our correction for
               the high-gravity component.\\
For the observations, we used the Atmospheric Dispersion Corrector
(ADC), and set up LRIS as follows.  A dichroic at 
5600\,\AA\ separated the incoming light into a blue and a red
channel.  For the blue channel we employed the $600{\rm\,lines\,mm^{-1}}$
grism, blazed at 4000\,\AA. The dispersion is 0.63\,\AA\, per pixel
and the spectral FWHM 4.1\,\AA\ (for the 1 arcsecond slit).  The
detectors consisted of two $4096\times 2048$ pixel Marconi CCDs
with each 15\,$\mu$m square pixel covering $0\farcs135$ on the side.
For the red channel we used a grating with $1200{\rm\,lines\,mm^{-1}}$
and blazed at 7500\,\AA.  The dispersion was 0.4\,\AA\ per pixel and
the spectral FWHM 2.1\,\AA\ (for the 1 arcsecond slit). The detector
for this channel was a mosaic of two LBNL CCDs, of the same size
and with the same plate scale as the blue ones (but read out binned
by 2 in the spatial direction).
}
\end{deluxetable}

\section{Spectroscopic Observations \&\ Data Reduction}
\label{sec:SpectroscopicObservations}

On the night of UT 2010 February 11, we undertook observations of
\SDSS\ with the Low Resolution Imaging Spectrograph (LRIS) mounted on
the Keck~I 10-m telescope \citep{occ+95}.  The night was clear and the
seeing typically about $0\farcs7$ (red camera) and $0\farcs8$ (blue
camera).  The observations are summarized in Table~\ref{tab:KeckLog}.
We started a sequence of observations of \SDSS\ using a $0\farcs7$
slit and later changed to $1\farcs0$ to better match the seeing.
Following this, we observed a number of internal light sources
(incandescent for flat fields, Hg/Kr/Zn/Ar and Ne/Ar lamps), as well as,
at a later time, observations of the flux standard Feige~34
\citep{o90} with the same spectrograph setup.

\begin{figure*}
   \centering
   \includegraphics[angle=-90,width=0.95\hsize]{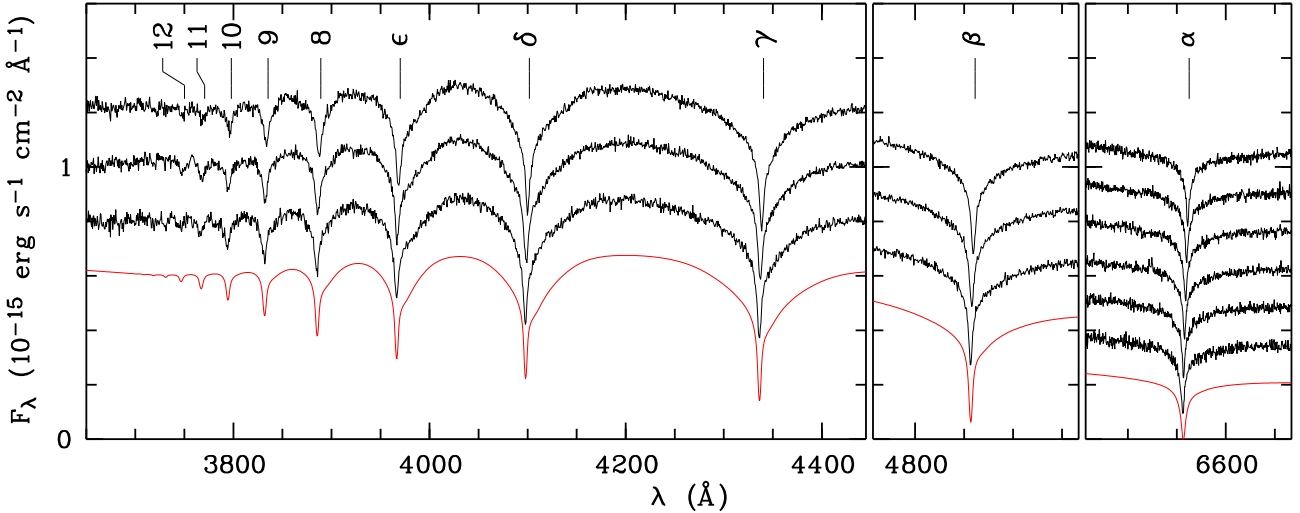} 
   \caption{The Balmer Series of \SDSS.  The spectra are ordered
     chronologically from bottom to top, and offset by 0.2 units (0.14
     units for H$\alpha$).  The second trace has offset 0.  The
     presence of Balmer lines up to H12 shows that one component of
     the binary has low gravity, while the wide wings on the lower
     Balmer lines suggest a high-gravity companion, with the asymmetry
     relative to the cores due to orbital motion and differences in
     gravitational redshift.  The lower, red curve shows the model fit
     described in the text, and shown in more detail in
     Figure~\ref{fig:SpectralFit}.}
   \label{fig:BalmerSeries}
\end{figure*}

For the reduction, we used the European Southern Observatory (ESO)
Munich Image Data Analysis System (MIDAS), and routines running in the
MIDAS environment.  For all images, we subtracted bias as determined
from the overscan regions.  For the blue images, we subsequently
corrected for small-scale variations in efficiency by dividing by a
spatially averaged flat field, normalized using a third-degree
polynomial, and with the bluest, poorly exposed part short-ward of
4000\,\AA\ replaced by unity.  For the red images, we simply divided
by the flat field, normalized using a bi-linear fit.  We extracted the
spectra using an optimal extraction procedure similar to that of
\citet{h86}, after subtracting sky determined from neighboring
regions.

Wavelength calibration was done using arc spectra.  For the blue
arm, fifth degree polynomial fits were required to give adequate
dispersion solutions, with typical root-mean-square residuals of
$0.02$\,\AA\ (for $14$ lines).  For the red arm, a fifth-degree
polynomial gave residuals of $\sim\!0.017\,$\AA\ (for $27$ lines).
We used the same wavelength solution for all spectra, but corrected
for possible drifts using the oxygen sky emission lines at 5577.340
and 6300.304\,\AA, in the blue and red, respectively.  For the blue,
the line is at the edge of our wavelength range, and hence our
wavelength scale may be slightly off.  Indeed, we find that the
velocities inferred from the blue spectra show a slit-size dependent
offset from those inferred from the red spectra; hence, we will use
the blue side only to (attempt to) measure relative velocities.

For flux calibration, we first corrected all spectra approximately for
atmospheric extinction using a curve made by combining the CFHT values
\citep*{bbd88} short-ward of 5200\,\AA\ with the better sampled La
Silla values long-ward of 5200\,\AA\ (ESO users manual 1993; see also
\citealt{tug77}).  Next, for the blue spectra, we calculated response
curves by comparing our observed spectra for Feige 34 with the
calibrated spectra of \citet{o90} as provided by
STScI:\footnote{ftp://ftp.stsci.edu/cdbs/calspec/} we slightly
smoothed our spectra to match the resolution, divided the two, and
smoothly interpolated the ratio.  Since all spectra were taken through
a relatively narrow slit, this should give good relative calibration,
but the absolute calibration may be off.  Folding our flux calibrated
spectra through the SDSS $u$, $g$, and $r$ response
curves,\footnote{http://www.sdss.org/dr6/instruments/imager/index.html\#filters}
we infer $g=16.6$, $u-g=0.41$, and $g-r=0.11$, which is in good
agreement with the values listed by \cite{bmt+09}: $16.844\pm0.004$,
$0.511\pm0.010$, and $0.112\pm0.007$, respectively, especially
considering that our spectra do not completely cover the u band.  For
plotting purposes, we rescaled our blue and red spectra to match the
$g$ and $r$, magnitudes, respectively.

\section{Spectral Analysis}
\label{sec:SpectralAnalysis}

The spectra show Balmer lines up to H12 and distinct asymmetries in
some of the Balmer absorption features, with the red wing shallower
than the blue one (Figure~\ref{fig:BalmerSeries}; note that there is
no sign of wiggles in the continuum such as those seen in Fig.~5 of
\citealt{bmt+09}).  This suggests that two sources contribute to the
Balmer features, with the second red-shifted relative to the one
responsible for the line cores.  The asymmetry is most pronounced in
H$\delta$ and H$\epsilon$, and less so for both lower and higher
Balmer lines.  The former suggests that the temperature of the second
component is higher than that of the most prominent one (but that the
emitting area is smaller), so that it contributes mostly at bluer
wavelengths, while the latter suggests its gravity must be higher, so
that the higher Balmer lines are less strong.

Granted that there are two sources contributing to the Balmer series
we were, nonetheless, puzzled by the absence of double line cores.
After some experimentation and contemplation we invoked rotational
broadening for the secondary star.

\begin{figure*}
   \centering
   \includegraphics[angle=-90,width=0.95\hsize]{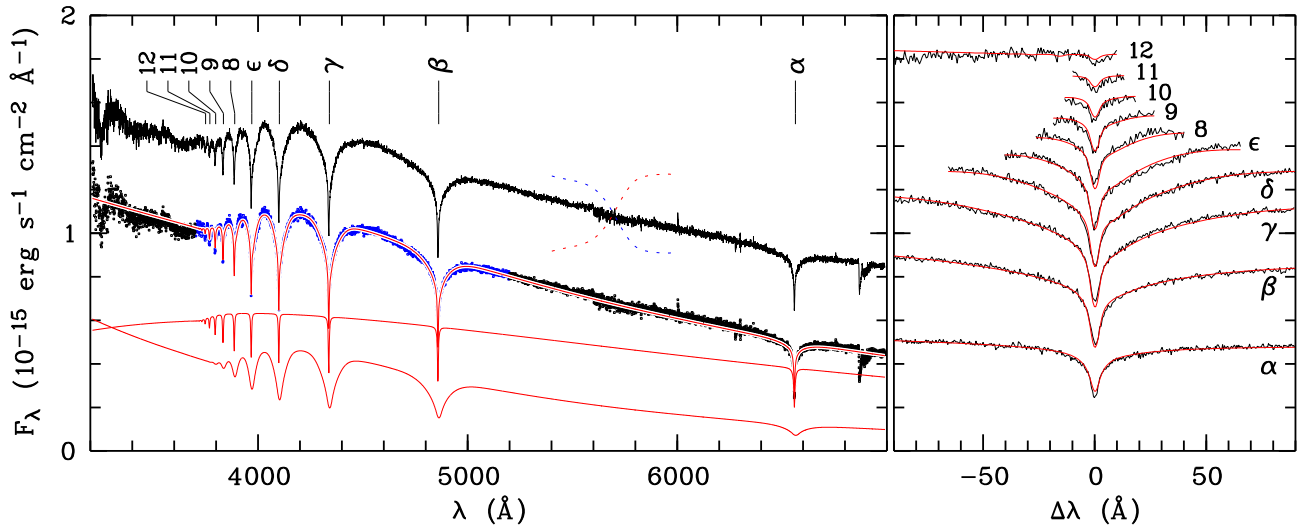}
   \caption{Spectrum of SDSS J1257+5428 and a model decomposition in
     two white-dwarf spectra.  In the left-hand panel, the top curve
     shows the observed spectrum, offset by 0.4 units (with the
     location of the dichroic indicated; we used the second set from
     Table~\ref{tab:KeckLog}, with two red spectra averaged).  The
     spectrum is repeated below, with the model overlaid (fit to the
     3700--5200\,\AA\ range [blue points]).  The two red lower curves
     show the two components in the model, one that has low gravity
     and relatively cool temperature, and one that has higher gravity
     and temperature, and is rotationally broadened.  In the
     right-hand panel, details of the fit around the Balmer lines are
     shown, with wavelength shifted to the rest frame of the
     low-gravity component, and offsets of 0.1 in flux added between
     all profiles beyond H$\beta$.}
\label{fig:SpectralFit}
\end{figure*}

\subsection{Model Fitting}

The above qualitative impressions are born out by fits with model
atmospheres.  For these fits, we modeled the blue spectra, in the
range of 3700--5200\,\AA\ (i.e., the range showing lines), with a
combination of two hydrogen model atmospheres, taken from a set kindly
provided by D.\ Koester.\footnote{This is Koester's most recent
  version of grid of white dwarf spectra. The basic model is described
  in \citet{kvn+09} and the latest version includes improved treatment
  of pressure broadening \citep{tb09}.  In detail, the grid spans
  $6000<T_{\rm eff}<30000\,$K in steps of 250 to 1000\,K at low and
  high temperature, and $6.0<\log g<9.5$ in steps of 0.25\,dex.}

We scanned a grid in effective temperature and surface gravity for
each component, as well as a set of rotational velocities $0<v_r\sin
i<2000{\rm\,km\,s^{-1}}$ for the higher gravity component (in steps of
$200{\rm\,km\,s^{-1}}$; we convolved the model using the analytical
profile of \citet{g05}, with a limb darkening coefficient of
0.3\footnote{The limb darkening coefficient is inferred from model
  specific intensities used previously to fit spectra of pulsating
  white dwarfs \citep{cvkw00}.  We ignored the variation in limb
  darkening over the lines.}).  For the lower-gravity component, we
took the smearing due to changing orbital velocity into account (using
the orbit of \citealt{mgs+10}; the largest effect is
$80{\rm\,km\,s^{-1}}$, substantially smaller than our resolution; the
motion of the high-gravity component during an integration is
negligible).

We accounted for the spectral resolution by convolving all models
with a truncated Gaussian (with width and truncation matching the
seeing and slit width, respectively), and resampling them at the
observed wavelengths.  At each grid point, we fitted for the two
best-fit velocities, the flux ratio, and a normalization (which,
to account for possible errors in our flux calibration, we allowed
to depend quadratically on wavelength).

\begin{deluxetable}{lll}
\tablewidth{\hsize}
\tablecaption{Spectroscopic Properties of the White Dwarfs in 
\SDSS\label{tab:BestFitParameters}}
\tablehead{
\colhead{Property}&\colhead{Phot.\ Primary}&\colhead{Phot.\ Secondary}
}
\startdata
\sidehead{Fit parameters}
$T_{\rm eff}~{\rm(K)}$ & $6250\pm250$ & $13000\pm500$ \\
$\log g~{\rm(cm\,s^{-2})}$ & $6.0\pm0.3$ & $8.5\pm0.2$ \\
$v_r\sin i~{\rm(km\,s^{-1})}$ & $<\!100$ & $800\pm300$ \\
$R_2/R_1$ & \multicolumn{2}{c}{$0.21\pm0.03$}\\
\sidehead{Derived parameters}
$M~(M_\odot)$ & $0.15\pm0.05$ & $0.92\pm0.13$\\
$R~(R_\odot)$ & $0.042\pm0.008$ & $0.0089\pm0.0012$
\enddata
\tablecomments{The parameters are for a model in which the binary is
  composed of two white dwarfs, with the secondary rotating rapidly.
  The limit to the rotation of the primary is derived separately from
  H$\alpha$; at the limit, the fit to the core is obviously worse.
  Note that the model fit is not formally acceptable (see text); the
  quoted uncertainties are what we believe are conservative estimates.}
\end{deluxetable}

Our fits yield a reasonably well-defined minimum, at the parameters
listed in Table~\ref{tab:BestFitParameters}.  We show the fit
overlaid on the whole spectrum in Fig.~\ref{fig:SpectralFit}.  One
sees that, overall, the fit is good, also outside the fitted range
(indeed, it reproduces the spectral slope remarkably well).

In detail, however, the fit has problems: the highest Balmer lines are
predicted slightly too weak, and some line cores are not reproduced
accurately (H$\alpha$ being not deep enough, and H$\gamma$ and H8 too
deep).  These effects might indicate that the true gravity of the
primary is slightly lower than that in our fit (which is at the lower
boundary of our grid).

Furthermore, the red wings of some of the lines are poorly matched:
that at H$\gamma$ is fainter than observed, and those at H$\epsilon$
and H8 are too bright.  We believe this difference is real rather
than due to, e.g., inaccurate flux calibration.  Indeed, in the
spectra of \citet{mgs+10}, the same discrepancies are seen.  One
possible reason, which we discuss further below (\S\ref{sec:Magnetism}),
is that the higher-gravity white dwarf is magnetic and that its
line profiles are (partially) broadened by Zeeman splitting.

Because of the above problems, even though the fit is hugely superior
to one using only a single white dwarf (cf.\ Figure~5 of
\citealt{bmt+09}), it is still not formally acceptable; for the spectra
taken through the 1\arcsec\ slit, we find $\chi^2_{\rm red}\simeq1.7$
(for 2408 data points and 13 parameters; $\chi^2_{\rm red}\simeq1.5$
for the single, less well-exposed spectrum taken through a $0\farcs7$
slit).  While the fit is internally consistent, in that the three blue
spectra give the same best-fit grid point, and that, e.g., fits to
H$\alpha$ and to higher Balmer lines give roughly consistent results
(for the latter, though, there is a much larger covariance with the
properties of the high-gravity component), we cannot know what offsets
are caused by the systematic differences.  We give what we believe are
conservative error estimates in Table~\ref{tab:BestFitParameters}.

Finally, we searched for the presence of
\ion{He}{1}~$\lambda5876$\,\AA\ (see \S\ref{sec:RevisitingMass}).
Inspecting the spectrum, we determined that a 2\% depression would be
clearly noticeable.  If an absorption feature existed at the limit of
instrumental resolution, then the equivalent width limit would be
$0.02\times 3\,$\AA=0.06\,\AA.  If the feature was broadened by
$800{\rm\,km\,s^{-1}}$, corresponding to $\Delta\lambda\sim\!15\,$\AA,
the equivalent width limit is $0.3\,$\AA.

\subsection{Comparison}

Comparing our inferred parameters with those given by \citet{mgs+10},
one finds qualitative agreement, in inferring a cool, low-gravity and
a hotter, high-gravity component, but what would appear significantly
different parameters (\citealt{mgs+10} find $T_{\rm
  eff,1}=7200\pm350$K, $\log g_1=6.85\pm0.10$, $T_{\rm
  eff,2}\simeq9800\pm1000\,$K, and $\log g_2\simeq9.0\pm0.4$).  It
appears that the most important difference is that in $T_{\rm eff,2}$:
if we fix this temperature, the remaining parameters become comparable
to those of \citet{mgs+10}, though the fit is much worse ($\chi^2_{\rm
  red}\simeq3.2$).  The same does not hold if we fix other parameters
(e.g., for $T_{\rm eff,1}=7250$, the fit quality is similarly poor for
a large range in $T_{\rm eff,2}$).  The underlying problem may be that
the Balmer lines have similar strength at 10,000\,K and 13,000\,K.
This leads to a degeneracy for normalized spectra such as those
presented by \citet{mgs+10}.  Usually, such degeneracies can be broken
by optical photometry, but here one can change the temperature of the
other component to compensate.  Fortunately, the ultraviolet data
presented by \citet{mgs+10} show conclusively that the
high-temperature solution is the correct one.

Turning now to the velocities, like \citet{mgs+10}, we find that our
fits yield precise measurements of the motion of the photometric
primary, in both our blue and red spectra, but not of that of the
secondary.  For the primary, for the blue spectra the accuracy is
limited by flexure in the spectrograph, which we appear to be able to
correct only partially using the 5577\,\AA\ sky line.  In
Table~\ref{tab:KeckLog}, we list the barycentric velocities inferred
from H$\alpha$; these are consistent with the radial-velocity
amplitude of $330{\rm\,km\,s^{-1}}$ and the systemic velocity of
$28.9{\rm\,km\,s^{-1}}$ found by \citet{bmt+09} and \citet{mgs+10}.
For the secondary, we find that while for every choice of parameters,
its velocity can be measured in the blue, and is consistently found
at a positive offset from the primary (as expected given the orbital
phases and its larger gravitational redshift), the precise values
depend strongly on the temperature of the primary, with changes of
250\,K leading to velocity differences of $\sim\!50$, 40, and
$30{\rm\,km\,s^{-1}}$ for the three different spectra.  Thus, we
cannot measure its motion.

\subsection{Inference}
\label{sec:Inference}

From the best-fit model parameters for the secondary
(Table~\ref{tab:BestFitParameters}) and using the cooling models of
\citet{wood92}, we deduce the following: $M_2=0.92\pm0.13\,M_\odot$
and $R_2=0.0089\pm0.0012\,R_\odot$, and a cooling age of
$\tau_c=0.7\pm0.3$\,Gyr.

For the primary, we start with the normalization in the spectral fits,
which yields the ratio of emitting areas for the two components, and
thus $R_2/R_1=0.21\pm0.03$.  Combined with the estimate for $R_2$, one
finds $R_1=0.042\pm0.008\,R_\odot$.  The spectral fits yield $\log
g_1=6.0\pm0.3$ (but more about this below).  The mass of the primary
can be directly computed from $\log g_1$ and the radius, $R_1$. The
mean value is $0.07\,M_\odot$; however, due to the large error in
$\log g$, the distribution of $M_1$ has a long tail with a variance
comparable to that of the mean.

Thus, we cannot determine the mass directly, but we can compare the
parameters with those for other low-mass white dwarfs.  The first were
seen as companions to millisecond pulsars (for a review,
\citealt{vkbjj05}) and more recently field objects were found via SDSS
\citep{kab+07}. Some notable objects are SDSS\,J0917+46 ($\log
g=5.48\pm 0.03$, $T_{\rm eff}=11288\pm 72\,$K; \citealt{kab+07});
NLTT\,11748 ($\log g=6.20\pm 0.05$, $T_{\rm eff}=8540\pm 50\,K$;
\citealt{kv09}); LP\,400-22 ($\log g=6.35\pm 0.05$, $T_{\rm eff}=
11170\pm 90\,$K; \citealt{vkv+09}). The latter is also notable because
it has a high space motion, even though its companion is another white
dwarf.

From evolutionary considerations, the true mass of the primary in
\SDSS\ is likely higher than the above estimate of $0.07\,M_\odot$
from $\log g$, since given its current orbital period, the progenitor
of must have evolved off the main sequence, which implies it must have
had a core mass of at least $\sim\!0.15\,M_\odot$.  This would imply a
true $\log g_1\gtrsim6.3$.  This is on the upper limit of our measurement,
but we note that the model atmospheres appear not to entirely reliable
in this low gravity regime. Specifically, models from different groups
gave results different by 0.4\,dex for the low-gravity, $\log
g\simeq6.5$ companion to \object{PSR J1012+5307} (see
\citealt{vkbjj05}).

\section{\SDSS, a double degenerate: ramifications}
\label{sec:DoubleDegenerateModel}

Above, we established that \SDSS\ is not a neutron star binary but a
{\it bona fide} double degenerate system with parameters given in
Table~\ref{tab:BestFitParameters}.  Here, we explore the consequences
of this conclusion.

\subsection{Distance \& Demography}
\label{sec:DistanceDemography}

The distance to \SDSS\ is increased because the flux of \SDSS\ is now
dominated by a lower mass white dwarf (with a larger radius and hence
a larger angular diameter relative to that in the \citealt{bmt+09}
model).  From our fit (see Table~\ref{tab:BestFitParameters}), we
infer a distance to \SDSS\ of about 140\,pc (instead of 50\,pc).

Is \SDSS\ unique?  There exists 129 systems (single white dwarfs,
white dwarf binaries and double degenerate binary systems) within
20\,pc \citep{sho+09}.  \citet{hso+08} find seven double degenerate
systems within 20\,pc.  Thus, within 140\,pc one expects to find about
2400 double degenerates.  A different estimate can be made by
combining the local density of white dwarfs of $4.5\times
10^{-3}{\rm\,pc^{-3}}$ with the finding of \citet{mm99} that about
1.7--19\% (95\% confidence level) of white dwarfs are double
degenerates.  This yields an expected 900--10000 double degenerates
within 140\,pc.  Thus, at first blush, \SDSS\ is not a rare system.

Furthermore, the systemic radial velocity of \SDSS\ is now
$\gamma_1-\gamma_E(1)$, where the former is the mean radial velocity
of the primary (see \S\ref{sec:RevisitingKinematics}) and
$\gamma_E(1)$ is the gravitational redshift for the primary. With the
revised mass, the redshift is small, $\gamma_E(1)=$2 to
$5{\rm\,km\,s^{-1}}$ (for $M_1=0.1$ to $0.2\,M_\odot$).  Thus, the 3-D
velocity of \SDSS\ is quite consistent with that of other double
degenerates.

The (exponential) scale height for the double degenerates, given
their measured velocity dispersion, is similar to that of (say) G
dwarfs, or about 350\,pc.

\subsection{Coalescence Time-scale}
\label{sec:CoalescenceTimeScale}

Double degenerates, especially in compact binaries, lose energy by
gravitational radiation and eventually coalesce.  The time for the
orbital period  to halve is (\citealt{lang80}, p. 600):
\begin{eqnarray}
	\frac{1}{P_{\rm b}}\frac{dP_{\rm b}}{dt} &=& -\frac{96}{5}\frac{G}{c^5}\frac{\mu M^2}{a^4}\cr
	 &=& \left(\frac{1}{2.8\times 10^7\,\mathrm{yr}}\right)
		\left(\frac{M}{M_\odot}\right)^{2/3}
		\left(\frac{\mu}{M_\odot}\right)
		\left(\frac{1\,\mathrm{hr}}{P_{\rm b}}\right)^{8/3}.
\end{eqnarray}
where $M=M_1+M_2$ and $\mu=M_1M_2/M$.  Integrating this equation to
zero period yields a coalescence time-scale of
\begin{equation}
	\tau_{\mathrm{GW}}=1.05\times 10^7\,\mathrm{yr}\;  
	\left(\frac{M}{M_\odot}\right)^{-2/3}
	\left(\frac{\mu}{M_\odot}\right)^{-1}
	\left(\frac{P_{\rm b}}{1\,\mathrm{hr}}\right)^{8/3}.
\end{equation}
%already updated
For $M_1=0.15\pm0.05\,M_\odot$ and $M_2=0.92\pm0.13\,M_\odot$ we
find $\tau_{\mathrm{GW}}=4.4\,$Gyr.  At the 1
percentile level the range is 2.5--15\,Gyr. Including the cooling
time-scale \citep{p91}  noted above (\S\ref{sec:Inference})
results in a total duration of $\sim\!5\,$Gyr.

\begin{figure}
   \centering
   \includegraphics[width=0.9\hsize]{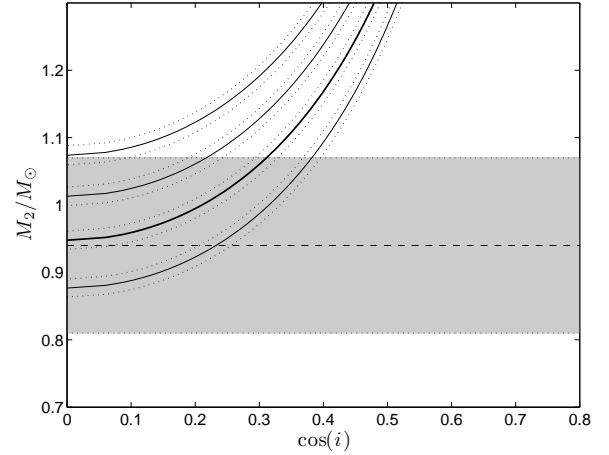}
   \caption[]{Mass of the photometric secondary as a function of $\cos
     i$ where $i$ is the angle between the orbital angular momentum
     and the line of sight (from bottom to top, $M_1=0.10, 0.15, 0.20,
     0.25 M_\odot$).  Here, we adopt the radial-velocity amplitude for
     the primary from \citet{mgs+10}, $K_1=330\pm
     2{\rm\,km\,s^{-1}}$.  The pair of dotted lines represent the
     uncertainty in $M_2$ (for a given value of $M_1$) arising solely
     from the 1.8\% uncertainty in the mass function, $\mathcal{M}_2$.
     The nominal value of $M_1= 0.15\,M_\odot$ is shown by the thick
     line.  The bounds on the secondary mass $M_2$ from spectral
     fitting is shown by the horizontal dotted lines. The totality of
     the data argue that $M_1\lesssim 0.2\,M_\odot$.}
   \label{fig:M2}
\end{figure}

\subsection{Merger Rate}

\citet{tks09}, accepting the masses of and distance to \SDSS\ of
\cite{bmt+09}, found a very large rate of coalescence of \SDSS-like
systems -- comparable to that of Ia supernovae.  The revision of the
distance, masses and scale height reduces the coalescence rate by a
large factor: using a scale height of 0.5\,kpc instead of 4\,kpc (see
\S\ref{sec:DistanceDemography}); a distance of 140\,pc to \SDSS\
instead of 50\,pc (\S\ref{sec:DistanceDemography}); and a lifetime of
$\sim\!5\,$Gyr instead of $2\,$Gyr (\S\ref{sec:CoalescenceTimeScale}),
our estimate of the Galactic coalescence rate is a factor $4/0.5\times
(140/50)^3\times 5/2$ smaller, or
\begin{equation} 
\gamma_{\mathrm{MW}} \sim 1\times 10^{-6}N\,{\rm yr^{-1}}, 
\label{eq:Rate} 
\end{equation}
where $N=1$ is the number of such systems detected (and consequently
suffers from severe Poisson uncertainty).

The rate given in Equation~\ref{eq:Rate} is not alarming
(corresponding to a mean time between events of $\sim\!1\,$Myr).
Furthermore, given that $M_1+M_2$ does not incontrovertibly exceed the
Chandrasekhar limit, the outcome of this coalescence is
(conservatively) not a type Ia supernova explosion, but rather the
formation an R~CrB helium giant or an AM CVn system (see the
discussion in \citealt{kba+09}).  The rate given in
Equation~\ref{eq:Rate} is small enough that, even if the combined mass
exceeds the Chandrasekhar limit, \SDSS-like systems are not major
contributors to the supernovae Ia rate.

\subsection{Broadening: Rotation?}

As noted in \S\ref{sec:SpectralAnalysis}, we are forced to invoke
broadening for the Balmer series of the secondary, and modeled this as
rotation.  We consider this simplest possibility first.  Our inferred
rotation rate is $v_r\sin i=800\pm300\,\mathrm{km\,s}^{-1}$ (see
Table~\ref{tab:BestFitParameters}), where $i$ is the inclination
between the rotation axis of the secondary and the line of sight.  We
will assume that the spin of the secondary is aligned with the angular
momentum of the binary.

For the entire range of plausible values for $M_2$ and the bounds for
$M_1$ we find that $\cos i<0.4$ and thus $\sin i>0.84$ (see
Figure~\ref{fig:M2}). In view of the errors on $v\sin i$, we will
approximate $\sin i\approx 1$.  Using our derived value for the radius
of the secondary, $R_2= 0.0089\pm0.0012\,R_\odot$ (see
\S\ref{sec:SpectralAnalysis}), we find that the rotational period,
$P\simeq 50^{+30}_{-15}\,$s, where the 30\% uncertainty in $v\sin i$
dominates the uncertainty.  We will adopt a round figure of 60\,s for
the rotation period of the secondary, and write corresponding angular
frequency as $\omega=2\pi/P$.

\subsubsection{Spin-up by Accretion}

The simplest possibility that comes to our mind as an explanation for
rapid rotation of the secondary is spin-up by accretion of matter.  In
order to lead to significant spin-up, this should happen in a
relatively long-lived phase, and hence it seems unlikely the system
was brought to its present state by a common-envelope phase
(furthermore, to survive a common-envelope phase, the system would
have to have been relatively wide, and one would expect a somewhat
more massive helium-core white dwarf).  Instead, it seems most likely
the evolution would be similar to what is invoked to explain the
millisecond periods of neutron stars in binaries for both long and
short periods ($P_{\rm b}>1\,$d, \citealt{wrs83}; $P_{\rm b}<1\,$d,
\citealt{ps88}).  Depending on the evolutionary status of the mass
donor and the orbital separation, mass transfer to a higher mass
accretor (in this case the white dwarf photometric secondary) is
driven by nuclear evolution, loss of angular momentum via stellar
winds (``magnetic braking'') or gravitational wave radiation.  The
typical accretion rate for short-period binaries is
$\sim\!10^{-9}\,M_\odot{\rm\,yr^{-1}}$, corresponding to an accretion
luminosity $GM_2\dot M/R_2\simeq10^{34}{\rm\,erg\,s^{-1}}$.

We first consider the case where the accreting white dwarf is
unmagnetized.  Its moment of inertia is $I_2=k_2M_2R_2^2$, with
$k_2=0.143$ appropriate for a $n=1.5$ polytrope \citep{bo55}.  For an
unmagnetized white dwarf, the accreted matter has the specific orbital
angular momentum appropriate for the radius of the white dwarf, or
$\Omega_2{R_2}^2$, where the orbital angular frequency
${\Omega_2}=(GM_2/{R_2}^3)^{1/2}\approx 7\,$s (for $M_2\approx
1\,M_\odot$).  Thus, for an accreted mass $\Delta M$, we have
\begin{eqnarray} 
	\Delta M \Omega_2{R_2}^2 &=& k_2M_2{R_2}^2\omega\cr 
	\frac{\Delta M}{M_2} &=&
	k_2\frac{\omega}{\Omega_2}=0.017.
\end{eqnarray}

It is empirically known that accreting white dwarfs in CVs do not
rotate rapidly (e.g., \citealt{gsh+05}), and one deduces that the
accreted angular momentum is lost, presumably in nova explosions.
This requires that for most CVs there is poor angular momentum
coupling between the accreted matter and the core \citep{lp98}.  Each
nova cycle, in effect, leaves both the mass and the spin state of the
accreting white dwarf unaffected.

For the low accretion rates one expects for the progenitor of \SDSS,
nova explosions will occur.  These happen when the accreted mass
exceeds a certain ``ignition'' value, the value of which depends on
the mass of the white dwarf.  According to \citet{tb04} the ignition
masses are $M_{\mathrm{ig}}\approx (6, 3, 2)\times 10^{-4}\,M_\odot$
for $(0.6,1.0,1.2)\,M_\odot$ white dwarfs. Thus, assuming that the
white dwarf accumulated no more than $M_{\mathrm{ig}}$ following the
very last nova explosion, the spin period can be no faster than about
an hour.

The minimum amount of accreted matter can be reduced for the case when
the accreting white dwarf is magnetized.  In this case, the accreting
matter has the specific angular momentum $\sqrt{GMr_A}$, where $r_A$
at the Alf\'ven radius.  This matter has more specific angular
momentum compared to the previous case by the factor $\sqrt{r_A/R_2}$,
which, once the white dwarf is spun up such that $r_A$ approaches the
corotation radius, scales as $\propto (\Omega_2/\omega)^{1/3}$ via
Kepler's third law.  Hence, in this case the minimum accreted matter
is reduced to
\begin{equation}
	\frac{\Delta M}{M_2}=k_2\frac{\omega}{\Omega_2}
	\left(\frac{\omega}{\Omega_2}\right)^{1/3} = 0.008\,M_\odot.
\end{equation}
However, even this mass exceeds the ignition masses by more than an
order of magnitude.  Thus, we reject the spin-up-by-accretion model,
at least for the case where the white dwarf is not magnetized or
magnetized sufficiently weakly that the accreted envelope does not
couple to the core.\footnote{It may be possible to find evolutionary
  histories in which mass transfer is much more rapid, or that have,
  e.g., an initial phase with rapid mass transfer, when the donor is
  still more massive than the white dwarf.  We did not investigate
  this further.}

\subsubsection{Intermediate Polar Model}

There exists a class of cataclysmic variables called intermediate
polars whose white dwarfs are found to be spinning quite rapidly.
These consist of a strongly magnetized white dwarf accreting matter
from the companion.  The most famous example is AE Aquarii with a spin
period of 32\,s. A number of intermediate polars have spin periods of
about a minute.\footnote{See
  http://asd.gsfc.nasa.gov/Koji.Mukai/iphome/iphome.html for a catalog
  of Intermediate Polars.} Motivated by the similarity of the spin
periods of intermediate polars and that inferred for \SDSS, we explore
an intermediate polar model for \SDSS.

Intermediate Polars undergo nova explosions (e.g., GK Per = Nova
Persei 1901; DQ Her = Nova Herculis 1934; see \citealt{l83}).  (Novae
also have been seen from polars, in which the white dwarf rotation is
magnetically locked to the orbit; e.g. V1500 Cyg = Nova Cygni 1975;
\citealt{ck88}.)  Thus, the rapid rotation seen in intermediate
polars, despite the novae, shows that the white dwarfs manage to
retain the accreted angular momentum.  We infer that the strong
magnetic field ensures rapid coupling between the accreted matter and
the rest of the white dwarf.  Subsequent novae explosions merely
result in ejection of the matter and a very modest fractional loss in
angular momentum, $\frac{2}{3}M_{\mathrm{ig}}/(k_2M_2)$ (where we
assumed $I_{\rm ej}=\frac{2}{3}M_{\rm ig}R_2^2$ for the ejected
shell).

Following the accretion of the minimum mass, the white dwarf will not
be spun up any further, but reach an equilibrium spin period, given by
\citep{gl79},
\begin{equation}
  P_{\mathrm{eq}}=6.3\times B_4^{6/7}L_{34}^{-3/7}
  \left(\frac{M}{M_\odot}\right) \left(\frac{R}{5\times10^8\,{\mathrm{cm}} }\right)\,{\mathrm{s}}.
\label{eq:Peq}
\end{equation}
Here, $B=10^4B_4\,$G is the dipole field of the white dwarf and
$L=10^{34}L_{34}\,{\rm erg\,s}^{-1}$ is the accretion luminosity
during the accretion phase.  With $M=0.9\,M_\odot$ and
$R=0.009\,R_\odot\simeq6\times 10^{8}\,$cm, we see that a modest
(dipolar) field of $1.3\times 10^5\,$G is sufficient to account for
the inferred period of $\sim\!60\,$s.

The dynamical equation for the spin-up is given by
\begin{equation}
	\frac{\dot\omega}{2\pi} \approx 3\times 10^{-16}\,\mathrm{Hz\,s}^{-1}\;
	L_{34}^{3/7}
	\mu_{30}^{2/7}
	I_{50}^{-1}
	\left(\frac{M}{M_\odot}\right)
	\left(\frac{R}{5\times 10^8\,{\mathrm{cm}}}\right)^{6/7},
	\label{eq:nudot}
\end{equation}
where $\mu=BR^3$ is the magnetic moment and the normalization is
that for a $10^4\,$G dipole field and radius appropriate for a solar
mass white dwarf.  The white dwarf would be spun up (assuming a
suitable accretion rate) in less than two million years.

We conclude that the intermediate polar model is a good explanation
for the origin of spin in \SDSS.  As an aside we make the following
observation.  The fastest intermediate polars have periods of about a
minute.  This then empirically suggests that a magnetic field strength
of $\sim\!10^5\,$G is the minimum field strength required to rapidly
couple the white dwarf and the accreted envelope.

\subsection{Broadening: Magnetism?}
\label{sec:Magnetism}

Above we assumed that the observed broadening in the high-gravity
spectrum is due to rotation.  An entirely different possibility is
that it is magnetic.  In a strong magnetic field, the Balmer series
are split into three components: two senses of circular polarization
and an unpolarized component.  To first order, the circular
polarization components are shifted by half the electron cyclotron
energy, or $0.0057{\rm\,eV}\;(B/10^6{\rm\,G})$, on each side of the
unperturbed unpolarized component.  At second order, all lines will
shift bluewards, with the shift substantially stronger for transitions
to higher excited states.

For the field strength of $10^5\,$G inferred above, the total expected
spread is $0.0011{\rm\,eV}$, which corresponds to a ``velocity width''
of about $200{\rm\,km\,s^{-1}}$ at H$\alpha$ (and smaller
proportionally with $\lambda^2$ for the higher Balmer lines). This in
itself will not have a noticeable effect on the model fits.

However, a possibility is that the white dwarf is not rotating rapidly
at all, but has a stronger field strength, say approaching $10^6\,$G.  In
this case, the observed broadening could be almost entirely due to
Zeeman splitting (e.g., as in WD~0637+477; \citealt{sss92}).  Indeed,
this might be the cause for the relatively poor spectral fits.  We
looked for the expected decrease in broadening and increase in
blueshift for the higher Balmer lines, but we were unable to
conclusively accept or reject the hypothesis that the Balmer lines are
broadened by Zeeman splitting.

\section{Testing the Model}

We found that \SDSS\ is composed of two white dwarfs, one relatively
cool one with low-gravity, and a higher-gravity, hotter one, whose
lines are substantially broadened.  A question that is left is what is
the cause for the broadening, rapid rotation with a period of about a
minute, or a strong, $\sim\!10^6\,$G, strength field.  The clearest way
to distinguish the two models would be by spectro-polarimetry.  For
the case of rapid rotation, this would also allow one to test whether,
as we contended, such rapid rotation requires the presence of a weaker
field, of $\sim\!10^5\,$G, similar to the fields inferred for
intermediate polars.  Rapid photometry (and spectrometry) could
perhaps even reveal the rotation period.  With $\log g=9.5$ and
$T_{\rm eff}=13,000\,$K, the secondary white dwarf lies close to the
ZZ Ceti strip \citep{gbf06} and so a search for pulsations could be
quite productive.
% If no evidence for moderate to strong magnetization
% is seen then we will have to conclude that the secondary star was born
% spinning rapidly or accreted at very high rate.

Apart from its intrinsic interest, \SDSS\ may also become a
useful test case for white dwarf models.  With sufficient sensitivity
and orbital coverage, it should be possible to measure velocity curves
for both components (i.e., derive $M_2/M_1$) and determine the
difference in gravitational redshift, (i.e., $(M/R)_2-(M/R)_1$).
Further model atmosphere fits would yield much more accurate
temperatures and gravities (i.e., $(M/R^2)_{1,2}$), and a precise
ratio of the emitting areas (i.e., $(R_2/R_1)^2$).  Combined with the
theoretical mass-radius relation, the system is thus strongly
overconstrained, and can be used to test various assumptions.  This
would be especially valuable for the low-mass companion, since, as
noted in \S\ref{sec:Inference}, for the very low-mass white dwarfs the
atmosphere and mass-radius relation are currently not as securely
established as is the case for regular white dwarfs.

\acknowledgments We thank E. S. Phinney, G. Nelemans, C. Badenes, and
L.\ Bildsten for discussion.  LRIS has been upgraded since its
commissioning and as a result is perhaps now the most efficient single
object optical spectrometer.  We thank the teams which made these
improvements possible (J.  McCarthy \&\ C. Steidel, leaders for
LRIS-Blue upgrade; C. Rokosi, leader for the LRIS-Red upgrade; and
J. Miller \&\ D. Phillips, leaders for the ADC sub-system).  We are
grateful to the staff of the WM Keck Observatory for their excellent
service and to the librarians who maintain the ADS and Simbad
databases.

\bibliographystyle{apj}
\bibliography{wd1257}

\begin{thebibliography}{50}
\expandafter\ifx\csname natexlab\endcsname\relax\def\natexlab#1{#1}\fi

\bibitem[{{Badenes} {et~al.}(2009){Badenes}, {Mullally}, {Thompson}, \&
  {Lupton}}]{bmt+09}
{Badenes}, C., {Mullally}, F., {Thompson}, S.~E., \& {Lupton}, R.~H. 2009,
  \apj, 707, 971

\bibitem[{{B{\`e}land} {et~al.}(1988){B{\`e}land}, {Boulade}, \&
  {Davidge}}]{bbd88}
{B{\`e}land}, S., {Boulade}, O., \& {Davidge}, T. 1988, Bulletin d'information
  du telescope Canada-France-Hawaii, 19, 16

\bibitem[{{Brisken} {et~al.}(2002){Brisken}, {Benson}, {Goss}, \&
  {Thorsett}}]{bbg+02}
{Brisken}, W.~F., {Benson}, J.~M., {Goss}, W.~M., \& {Thorsett}, S.~E. 2002,
  \apj, 571, 906

\bibitem[{{Brooker} \& {Olle}(1955)}]{bo55}
{Brooker}, R.~A. \& {Olle}, T.~W. 1955, \mnras, 115, 101

\bibitem[{{Chlebowski} \& {Kaluzny}(1988)}]{ck88}
{Chlebowski}, T. \& {Kaluzny}, J. 1988, Acta Astronomica, 38, 329

\bibitem[{{Clemens} {et~al.}(2000){Clemens}, {van Kerkwijk}, \& {Wu}}]{cvkw00}
{Clemens}, J.~C., {van Kerkwijk}, M.~H., \& {Wu}, Y. 2000, \mnras, 314, 220

\bibitem[{{Cordes} \& {Chernoff}(1997)}]{cc97}
{Cordes}, J.~M. \& {Chernoff}, D.~F. 1997, \apj, 482, 971

\bibitem[{{Deller} {et~al.}(2009){Deller}, {Tingay}, {Bailes}, \&
  {Reynolds}}]{dtb+09}
{Deller}, A.~T., {Tingay}, S.~J., {Bailes}, M., \& {Reynolds}, J.~E. 2009,
  \apj, 701, 1243

\bibitem[{{Deller} {et~al.}(2008){Deller}, {Verbiest}, {Tingay}, \&
  {Bailes}}]{dvt+08}
{Deller}, A.~T., {Verbiest}, J.~P.~W., {Tingay}, S.~J., \& {Bailes}, M. 2008,
  \apjl, 685, L67

\bibitem[{{Dodson} {et~al.}(2003){Dodson}, {Legge}, {Reynolds}, \&
  {McCulloch}}]{dlr+03}
{Dodson}, R., {Legge}, D., {Reynolds}, J.~E., \& {McCulloch}, P.~M. 2003, \apj,
  596, 1137

\bibitem[{{Faherty} {et~al.}(2007){Faherty}, {Walter}, \& {Anderson}}]{fwa07}
{Faherty}, J., {Walter}, F.~M., \& {Anderson}, J. 2007, \apss, 308, 225

\bibitem[{{G{\"a}nsicke} {et~al.}(2005){G{\"a}nsicke}, {Szkody}, {Howell}, \&
  {Sion}}]{gsh+05}
{G{\"a}nsicke}, B.~T., {Szkody}, P., {Howell}, S.~B., \& {Sion}, E.~M. 2005,
  \apj, 629, 451

\bibitem[{{Ghosh} \& {Lamb}(1979)}]{gl79}
{Ghosh}, P. \& {Lamb}, F.~K. 1979, \apj, 234, 296

\bibitem[{{Gianninas} {et~al.}(2006){Gianninas}, {Bergeron}, \&
  {Fontaine}}]{gbf06}
{Gianninas}, A., {Bergeron}, P., \& {Fontaine}, G. 2006, \aj, 132, 831

\bibitem[{{Gray}(2005)}]{g05}
{Gray}, D.~F. 2005, {The Observation and Analysis of Stellar Photospheres}
  (Cambridge University: Cambridge University Press)

\bibitem[{{Hamada} \& {Salpeter}(1961)}]{hs61}
{Hamada}, T. \& {Salpeter}, E.~E. 1961, \apj, 134, 683

\bibitem[{{Holberg} {et~al.}(2008){Holberg}, {Sion}, {Oswalt}, {McCook},
  {Foran}, \& {Subasavage}}]{hso+08}
{Holberg}, J.~B., {Sion}, E.~M., {Oswalt}, T., {McCook}, G.~P., {Foran}, S., \&
  {Subasavage}, J.~P. 2008, \aj, 135, 1225

\bibitem[{{Horne}(1986)}]{h86}
{Horne}, K. 1986, \pasp, 98, 609

\bibitem[{{Kawka} \& {Vennes}(2009)}]{kv09}
{Kawka}, A. \& {Vennes}, S. 2009, \aap, 506, L25

\bibitem[{{Kilic} {et~al.}(2007){Kilic}, {Allende Prieto}, {Brown}, \&
  {Koester}}]{kab+07}
{Kilic}, M., {Allende Prieto}, C., {Brown}, W.~R., \& {Koester}, D. 2007, \apj,
  660, 1451

\bibitem[{{Kilic} {et~al.}(2009){Kilic}, {Brown}, {Allende Prieto}, \&
  {Kenyon}}]{kba+09}
{Kilic}, M., {Brown}, W.~R., {Allende Prieto}, C., \& {Kenyon}, S.~J. 2009,
  arXiv:0911.1781

\bibitem[{{Koester} {et~al.}(2009{\natexlab{a}}){Koester}, {Kepler},
  {Kleinman}, \& {Nitta}}]{kkk+09}
{Koester}, D., {Kepler}, S.~O., {Kleinman}, S.~J., \& {Nitta}, A.
  2009{\natexlab{a}}, Journal of Physics Conference Series, 172, 012006

\bibitem[{{Koester} {et~al.}(2009{\natexlab{b}}){Koester}, {Voss},
  {Napiwotzki}, {Christlieb}, {Homeier}, {Lisker}, {Reimers}, \&
  {Heber}}]{kvn+09}
{Koester}, D., {Voss}, B., {Napiwotzki}, R., {Christlieb}, N., {Homeier}, D.,
  {Lisker}, T., {Reimers}, D., \& {Heber}, U. 2009{\natexlab{b}}, \aap, 505,
  441

\bibitem[{{Lang}(1980)}]{lang80}
{Lang}, K.~R. 1980, {Astrophysical Formulae. A Compendium for the Physicist and
  Astrophysicist.} (Berlin, Germany: Springer Verlag)

\bibitem[{{Livio}(1983)}]{l83}
{Livio}, M. 1983, \aap, 121, L7

\bibitem[{{Livio} \& {Pringle}(1998)}]{lp98}
{Livio}, M. \& {Pringle}, J.~E. 1998, \apj, 505, 339

\bibitem[{{Lyne} {et~al.}(1998){Lyne}, {Manchester}, {Lorimer}, {Bailes},
  {D'Amico}, {Tauris}, {Johnston}, {Bell}, \& {Nicastro}}]{lml+98}
{Lyne}, A.~G., {Manchester}, R.~N., {Lorimer}, D.~R., {Bailes}, M., {D'Amico},
  N., {Tauris}, T.~M., {Johnston}, S., {Bell}, J.~F., \& {Nicastro}, L. 1998,
  \mnras, 295, 743

\bibitem[{{Marsh} {et~al.}(2010){Marsh}, {Gaensicke}, {Steeghs}, {Southworth},
  {Koester}, {Harris}, \& {Merry}}]{mgs+10}
{Marsh}, T.~R., {Gaensicke}, B.~T., {Steeghs}, D., {Southworth}, J., {Koester},
  D., {Harris}, V., \& {Merry}, L. 2010, arXiv:1002.4677

\bibitem[{{Maxted} \& {Marsh}(1999)}]{mm99}
{Maxted}, P.~F.~L. \& {Marsh}, T.~R. 1999, \mnras, 307, 122

\bibitem[{{Mullally} {et~al.}(2009){Mullally}, {Badenes}, {Thompson}, \&
  {Lupton}}]{mbt+09}
{Mullally}, F., {Badenes}, C., {Thompson}, S.~E., \& {Lupton}, R. 2009, \apjl,
  707, L51

\bibitem[{{Nauenberg}(1972)}]{n72}
{Nauenberg}, M. 1972, \apj, 175, 417

\bibitem[{{Nelemans} {et~al.}(2005){Nelemans}, {Napiwotzki}, {Karl}, {Marsh},
  {Voss}, {Roelofs}, {Izzard}, {Montgomery}, {Reerink}, {Christlieb}, \&
  {Reimers}}]{nnk+05}
{Nelemans}, G., {Napiwotzki}, R., {Karl}, C., {Marsh}, T.~R., {Voss}, B.,
  {Roelofs}, G., {Izzard}, R.~G., {Montgomery}, M., {Reerink}, T.,
  {Christlieb}, N., \& {Reimers}, D. 2005, \aap, 440, 1087

\bibitem[{{Oke}(1990)}]{o90}
{Oke}, J.~B. 1990, \aj, 99, 1621

\bibitem[{{Oke} {et~al.}(1995){Oke}, {Cohen}, {Carr}, {Cromer}, {Dingizian},
  {Harris}, {Labrecque}, {Lucinio}, {Schaal}, {Epps}, \& {Miller}}]{occ+95}
{Oke}, J.~B., {Cohen}, J.~G., {Carr}, M., {Cromer}, J., {Dingizian}, A.,
  {Harris}, F.~H., {Labrecque}, S., {Lucinio}, R., {Schaal}, W., {Epps}, H., \&
  {Miller}, J. 1995, \pasp, 107, 375

\bibitem[{{Phinney}(1991)}]{p91}
{Phinney}, E.~S. 1991, \apjl, 380, L17

\bibitem[{{Pylyser} \& {Savonije}(1988)}]{ps88}
{Pylyser}, E. \& {Savonije}, G.~J. 1988, \aap, 191, 57

\bibitem[{{Schmidt} {et~al.}(1992){Schmidt}, {Stockman}, \& {Smith}}]{sss92}
{Schmidt}, G.~D., {Stockman}, H.~S., \& {Smith}, P.~S. 1992, \apjl, 398, L57

\bibitem[{{Sion} {et~al.}(2009){Sion}, {Holberg}, {Oswalt}, {McCook}, \&
  {Wasatonic}}]{sho+09}
{Sion}, E.~M., {Holberg}, J.~B., {Oswalt}, T.~D., {McCook}, G.~P., \&
  {Wasatonic}, R. 2009, \aj, 138, 1681

\bibitem[{{Thompson} {et~al.}(2009){Thompson}, {Kistler}, \& {Stanek}}]{tks09}
{Thompson}, T.~A., {Kistler}, M.~D., \& {Stanek}, K.~Z. 2009, ArXiv:0912.0009

\bibitem[{{Townsley} \& {Bildsten}(2004)}]{tb04}
{Townsley}, D.~M. \& {Bildsten}, L. 2004, \apj, 600, 390

\bibitem[{{Tremblay} \& {Bergeron}(2009)}]{tb09}
{Tremblay}, P. \& {Bergeron}, P. 2009, \apj, 696, 1755

\bibitem[{{Tremblay} {et~al.}(2009){Tremblay}, {Bergeron}, \& {Dupuis}}]{tbd09}
{Tremblay}, P., {Bergeron}, P., \& {Dupuis}, J. 2009, Journal of Physics
  Conference Series, 172, 012046

\bibitem[{{Tremblay} {et~al.}(2010){Tremblay}, {Bergeron}, {Kalirai}, \&
  {Gianninas}}]{tbk+10}
{Tremblay}, P., {Bergeron}, P., {Kalirai}, J.~S., \& {Gianninas}, A. 2010,
  ArXiv:1002.3585

\bibitem[{{Tug}(1977)}]{tug77}
{Tug}, H. 1977, The Messenger, 11, 7

\bibitem[{{van Kerkwijk} {et~al.}(2005){van Kerkwijk}, {Bassa}, {Jacoby}, \&
  {Jonker}}]{vkbjj05}
{van Kerkwijk}, M.~H., {Bassa}, C.~G., {Jacoby}, B.~A., \& {Jonker}, P.~G.
  2005, in Astronomical Society of the Pacific Conference Series, Vol. 328,
  Binary Radio Pulsars, ed. {F.~A.~Rasio \& I.~H.~Stairs}, 357

\bibitem[{{van Kerkwijk} \& {Kaplan}(2007)}]{vkk07}
{van Kerkwijk}, M.~H. \& {Kaplan}, D.~L. 2007, \apss, 308, 191

\bibitem[{{Vennes} {et~al.}(2009){Vennes}, {Kawka}, {Vaccaro}, \&
  {Silvestri}}]{vkv+09}
{Vennes}, S., {Kawka}, A., {Vaccaro}, T.~R., \& {Silvestri}, N.~M. 2009, \aap,
  507, 1613

\bibitem[{{Webbink} {et~al.}(1983){Webbink}, {Rappaport}, \&
  {Savonije}}]{wrs83}
{Webbink}, R.~F., {Rappaport}, S., \& {Savonije}, G.~J. 1983, \apj, 270, 678

\bibitem[{{Wood}(1992)}]{wood92}
{Wood}, M.~A. 1992, \apj, 386, 539

\bibitem[{{York} {et~al.}(2000)}]{yaa+00}
{York}, D.~G. {et~al.} 2000, \aj, 120, 1579

\end{thebibliography}
\end{document}